# AQUEOUS ALTERATION ON ASTEROIDS SIMPLIFIES SOLUBLE ORGANIC MATTER MIXTURES

JUNKO ISA[1,2,3], FRANÇOIS-RÉGIS ORTHOUS-DAUNAY[3], PIERRE BECK[3], CHRISTOPHER D. K. HERD[4], VERONIQUE VUITTON[3], and LAURÈNE FLANDINET[3]

ABSTRACT

Biologically relevant abiotic extraterrestrial soluble organic matter (SOM) has been widely investigated to study the origin of life and the chemical evolution of protoplanetary disks. Synthesis of biologically relevant organics, in particular, seems to require aqueous environments in the early solar system. However, SOM in primitive meteorites includes numerous chemical species besides the biologically relevant ones and the reaction mechanisms that comprehensively explain the complex nature of SOM are unknown. Besides, the initial reactants, which formed before asteroid accretion, were uncharacterized. We examined the mass-distribution of SOM extracted from three distinct Tagish Lake meteorite fragments, which exhibit different degrees of aqueous alteration though they originated from a single asteroid. We report that mass-distributions of SOM in the primordial fragments are well fit by the Schulz-Zimm (SZ) model for the molecular weight distribution patterns found in chain-growth polymerization experiments. Also, the distribution patterns diverge further from SZ with increasing degrees of aqueous alteration. These observations imply that the complex nature of the primordial SOM 1) was established before severe alteration on the asteroid, 2) possibly existed before parent body accretion, and 3) later became simplified on the asteroid. Therefore, aqueous reactions on asteroids are not required conditions for cultivating complex SOM. Furthermore, we found that overall H/C ratios of SOM decrease with increasing of aqueous alteration, and the estimate of H loss from the SOM is 10-30%. Organics seem to be a significant $H_2$ source that may have caused subsequent chemical reactions in the Tagish Lake meteorite parent body.

## 1. INTRODUCTION

[1] Corresponding author junko.isa@univ-grenoble-alpes.fr
[2] Earth-Life Science Institute, Tokyo Institute of Technology, Tokyo, Japan
[3] Institut de Planétologie et Astrophysique de Grenoble, Université Grenoble Alpes, Grenoble, France
[4] Department of Earth and Atmospheric Sciences, University of Alberta, Edmonton, AB, Canada

Abiotic organic compounds are common and widely distributed in solar system objects (Schmitt-Kopplin et al. 2010). Understanding their formation mechanism requires accurate knowledge of their composition at the time of planetesimal accretion, 4.5 Gy ago, which is complexified by geological processes that occurred on these small bodies over the history of the solar system. In particular, the low-temperature alteration of asteroids, which normally took place a few My after accretion at temperatures ranging from ~0℃ to ~150℃ under aqueous conditions (Suttle et al. 2021) is considered to be an important driver of chemical evolution. This process may have played an essential geological role for producing the building blocks of life, for example, through amplification of L-isovaline excesses (Glavin & Dworkin 2009). A challenge of meteoritic organics studies is to distinguish the pre-accretionary (hereafter "primordial") features from the secondary components that are produced by geological processes on their original asteroidal parent bodies. Inorganic components are distinct across the different chondrites classes and groups: examples include refractory abundances (e.g., Wasson & Kallemeyn 1988), oxidation states (e.g., Urey & Craig 1953), bulk stable-isotope compositions (e.g., Warren 2011), and the size distribution of chondrules (e.g., Jones 2012). On top of the variation of the initial materials, the degree of aqueous alteration and thermal metamorphism can vary for a given meteorite group.

In our study, we examine the Tagish Lake carbonaceous chondrite, an organic rich (~2.5 wt.% organic C), C2 ungrouped meteorite (Grady et al. 2002) that fell on the frozen lake on January 18$^{th}$ in 2000 in northwestern British Columbia, Canada. After the fall, several hundred grams were recovered on the 25$^{th}$ and 26$^{th}$ of January and these meteorite fragments that had minimum exposure to terrestrial organics are so called "pristine" samples. Later, in the spring 2000, more meteorite fragments were recovered and these are so called "degraded" samples because of terrestrial water contamination (Brown et al. 2000). That pristine collection has provided us a unique opportunity to study abiotic organics and the effects of aqueous alteration on its parent body (e.g., Herd et al. 2011; Gilmour et al. 2019; Simkus et al. 2019). The Tagish Lake meteorite is comprised of several lithologies that experienced different degrees of secondary processing on the same parent body (Zolensky et al. 2002; Herd et al. 2011). We selected three different previously well-studied lithologies: TL5b, TL11h, and TL11i. As a part of the rock, the organics experienced the same geological processes as adjoining inorganic minerals whose textures recorded the degrees of aqueous alteration. Observed trends within the organic matter in Tagish Lake have previously been tied to the degree of aqueous alteration, in the order TL5b (the least altered) < TL11h < TL11i (Herd et al. 2011; Blinova et al. 2014a; Gilmour et al. 2019). The TL5b lithology has the highest abundance of chondrules and lowest abundance of phyllosilicate material; TL11h and TL11i lithologies have increasing proportions, respectively, of parent body alteration products, including pore-filling clays (Blinova et al. 2014a). Using thermogravimetric analysis and infrared transmission spectroscopy, Gilmour et al. (2019) showed that these (and other) Tagish Lake lithologies are comparable to the least-altered members of the CM chondrites, equivalent to petrologic types 2.0 (TL5b) through 1.6 (TL11i).

A previous study of insoluble organic matter (IOM) present in Tagish Lake revealed that modification due to the aqueous alteration on the parent body positively and linearly correlates with $\delta D$ values and H/C ratios, which have been interpreted to reflect the effects of varying degrees of aqueous alteration. For example, the IOM in TL5b (the least altered) have the highest H/C ratios and heavy $\delta D$ values among the samples from the pristine Tagish Lake collection (Figure 1(b) in Alexander et al. 2014). Thus, one could imagine that the vast majority of soluble organic matter was also subject to reactions that lead to synthesis or degradation due to their nature to be reactive; indeed, the SOM in Tagish Lake shows similar relationships (Herd et al. 2011; Hilts et al. 2014). Even if the alteration effect does not seem to modify the structure of the macromolecules in the IOM greatly, it may have a dramatic impact on smaller molecules comprising the SOM, either producing them or significantly changing their structure. In particular, the SOM in carbonaceous chondrites is of significant interest, especially the presence of potential biologically relevant compounds such amino acids, sugars, and fatty acids. In Tagish Lake, the most prominent bulk SOM includes monocarboxylic acids (MCAs) and amino acids that are heterogeneously distributed (Hilts et al. 2014). However, it is still unclear concerning what pre-accretionary precursors were present and which formation processes occurred. Thus, it is ambiguous how the organics present in chondrites were formed from those pre-accretionary materials. To identify evidence of aqueous alteration of SOM on asteroidal parent bodies, we obtained high-resolution mass spectra of three types of lithologies found in Tagish Lake that experienced different levels of aqueous alteration. We applied a holistic approach to the size distribution of organic molecules. First, we assessed the size diversity of SOM molecular distributions by using the number of heteroatoms, and Double Bond Equivalent values (DBE = C+1-H/2+N/2) to estimate the level of unsaturation. Second, we fit the data by using a known polymer synthesis model. These data allow one to test previously suggested reaction pathways of organic matter formation in carbonaceous chondrites such as a formose-like reaction (e.g., Furukawa et al. 2019).

## 2. METHOD

Approximately 10mg of powdered samples were extracted by using a methanol and toluene (1:2) mixture. The meteorite extract and solvent were separated from the meteorite powder by centrifuging. The direct infusion technique was used with a high-resolution mass spectrometer, Thermo LTQ Orbitrap XL instrument coupled with an electrospray ionization (ESI) source, in the m/z range of 150—800. The positive ions were analyzed with a resolving power of $m/\Delta m$ ~ 100000 at m/z = 400u.

## 3. RESULTS

We found that the structures in the individual mass-spectrum from the three extracts from 5b, 11h, 11i are distinct in terms of their size distributions and relative DBE values. From one sample to another, the molecular size distribution narrows as the degree of aqueous alteration increases. The mass spectra of TL5b can be split into several sub spectra in which the DBE and the number of O and N atoms are invariant for the sake of clarity. The shape of the envelope of these sub spectra, plotted in Fig 1 appears to be similar regardless of a number of heteroatoms or DBE value. With CHNO only involved in this study, sub spectra turn out to be individual $CH_2$ families and they indeed are well fit by the Schulz-Zimm (SZ) distribution that models the molecular weight distribution of chain-growth polymerization experiments (Scholz 1939; Zimm 1948), see Fig 2. Overall goodness of fit ($R^2$) for the individual $CH_2$ family (number of family members >10) are shown in S3. Notably, the distribution pattern becomes asymmetric and sharp with increasing degree of aqueous alteration and the most intense peak of a $CH_2$ family decreases in mass with increasing degree of aqueous alteration. These deviations from the SZ distribution seen in the mass spectra are associated with the degree of aqueous alteration. The spiked and tailing from a SZ distribution indicates that the size diversity is reduced in a $CH_2$ family as shown by entropy values calculated for a given $CH_2$ family (Fig 2D).
The results indicate that the hydrogen content in the compounds decreases in more altered samples. While the SOM mass-range among the samples are unchanged, the maximum DBE increases (Fig 3). The same figure shows that the least altered sample is the only one that has heavy molecules with low DBE (yellow). The altered samples systematically populate the highest DBE domains (blue). This trend indicates that the carbon atoms in these compounds became oxidized. This trend is also observed in the chemical formulae that are common among the three spectra (black in Fig 3). Normalized intensities among the three samples indicate that the larger DBE intensity increases with higher degree of alteration given an equal number of carbons in the formulae. Furthermore, we can see that the relatively small molecule intensities are higher in the most altered sample for the same $CH_2$ family S5.

## 4. DISCUSSION

It is shown in Fig 1 that a high degree of similarity exists in terms of pattern regardless of the number of N or O atoms in the chemical formulae. The least altered sample is the one where the SZ model explains the observations the best for each individual family and with the best repeatability among the sample family set. These qualities make the SZ model a good candidate to interpret the growth mechanism responsible for the $CH_2$ variability observed in the mass spectrum. The applicability of the SZ model to different families is remarkable because these patterns and their stability are unlikely to result from reactions caused by a particular functional group (i.e., heteroatoms). It is natural to assume that heteroatoms, which can polarize molecules, should have played a significant role if polymerization occurred in an aqueous fluid in the presence of inorganic ions on the parent body. Therefore, if $CH_2$ chains grew during aqueous alteration, one would expect distinct $CH_2$ polymerization patterns depending on the type and number of heteroatoms. On the contrary, the ubiquity of the SZ patterns tends to show that the complex organic mixture has been synthesised by homogeneous reactions allowing a high degree of branching carbon chains (Pladis & Kiparissides 1998).

The presence of the SZ pattern provides insights into the formation pathways of the organic compounds in the least altered Tagish Lake sample. Many chemical reaction pathways invoked in previous studies to explain chondrite organics cannot explain the SZ pattern found in the primordial $CH_2$ polymerization. In particular, condensation reactions in aqueous environments on the parent body that have been proposed to account for IOM-like residues (Cody et al. 2011) or alkylated homologs of N-bearing cyclic compounds (Naraoka et al. 2017) are not capable of explaining the observed SOM distribution. In general, step-growth polymerization involves reactions between two functional groups provided by the monomers and generate molecular weight variability well described by so-called Anderson-Flory-Schultz distributions (Förtsch et al. 2015) (see the best fit S6). Such distributions can be found in Fischer-Tropsch type synthesized mixtures and abiotic RNA polymerization (Spaeth & Hargrave 2020), etc. Therefore, the step-growth hypothesis can be rejected in favour of chain-growth polymerization to explain the $CH_2$ family groups observed in the least altered sample.

For the altered samples where a significant deviation from the SZ distribution is found, two hypotheses can be invoked to explain the difference with more primordial samples. One may argue that their occurrence in the most altered part of Tagish Lake is fortuitous and they are simply a mixture of several homologous series from primordial organic compounds. So, one simple polymerization profile cannot describe them. This possibility cannot be strictly ruled out. However, three remarks should be noted. 1) The observed SOM represents the bulk compositions. It is shown that the bulk elemental compositions of the three lithologies are relatively isochemical (Blinova et al. 2014b) (S7). Thus, sample heterogeneity caused by chondrite accretion processes for the inorganic components are likely less pronounced. 2) Although the Tagish Lake meteorite contains foreign clasts, the compositions of the individual specimens TL5b, TL11h, and TL11i appear to be consistent and can thus be described as lithologies (Blinova et al. 2014a). 3) The distortion from a SZ distribution is gradually enhanced with increasing degree of alteration. Such an inclination is not expected if the trend was caused by a simple mixture of multiple primordial organic components. Indeed, even chromatographic separation would not be able to distinguish molecules resulting from several synthesis processes in the case that they are identical.

The alternative, which is favoured here according to Occam's razor principle, is that secondary processes modified an initial synthesis pattern. Although TL5b is the least altered sample, it is an altered rock (petrologically classified as type 2). Therefore, it is not surprising to see evidence of alteration effects on SOM in our result. A previous study located amorphous silicates in the lithology TL5b matrix, which was interpreted to infer a relatively low alteration temperature $\leqq 50$ °C (Blinova et al. 2014a). We can recognize these mild alteration effects in our measured SOM distribution. Although most $CH_2$ families in the least altered sample, TL5b, maintained their SZ profile throughout the alteration, as depicted by data points with higher entropy and $R^2$ values close to 1 (Fig 2D), some distortions from the SZ profile are found in a few $CH_2$ families. That alteration evidence is depicted as the yellow data points with lower entropy and smaller $R^2$ values (Fig 2D). Therefore, the SZ pattern extrapolated from the three samples can be taken as representative of the original soluble organic makeup that was perhaps established before accretion of the Tagish Lake parent body. The possible chemical reaction field to establish the SZ profile is not limited to aqueous conditions or low temperature. For example, the polymers can be synthesized from gas by chain-growth reactions (Alves et al. 2021). And, high temperature gas phase experiments can form various organics ranging from IOM-like material to small molecules comparable to SOM, as well as the production of the Xenon mass fraction observed in the matrix known as the Q-phase (Kuga et al. 2015; Kuga et al. 2017; Bekaert et al. 2018). The SOM mass distributions of those materials indeed can be well fit by the SZ model (S8). We show the distributions and the SZ fit in a figure together with the data (S9). These high-temperature reactions provide an interesting pathway to form SOM. High-temperature conditions can be found in irradiated parts of the solar nebula. Extra-terrestrial samples also record massive high-temperature processing of protoplanetary disk materials by the widespread occurrence of chondrules. These formed under high temperature and high ambient gas pressure transient events (by evaporating the gas during their formation) in the solar nebula (Alexander & Ebel 2012) and references therein. We do not rule out the possibility of the formation mechanisms besides in a vapor phase. And yet, forming the SZ profile of the SOM complex before accretion is plausible and consistent with the rest of the chondrite components.

Aqueous alteration processes resulted in an oxidized signature (that is, the loss of H) of the SOM, which is consistent with the previously observed linear correlation in H/C ratios and δD of Tagish Lake IOM (Cody & Alexander 2005; Alexander et al. 2007; Herd et al. 2011; Alexander et al. 2014) and inferences from previous IOM and SOM studies (Hilts et al. 2014; Quirico et al. 2018). The present measurement shows that the larger molecules were spared from removal throughout the alteration (Fig 3) as depicted in the point clouds. A chemical change is observed in the composition of the individual molecules: the most altered samples shift towards higher DBE. Such hydrogen elimination reactions can take place by adding energy to a system. The H-loss from organics was confirmed after heating meteorite IOM (Yabuta et al. 2007; Oba & Naraoka 2009) and after irradiating meteorite SOM (Orthous-Daunay et al. 2019). In the reactions that occurred during aqueous alteration, the processes were likely facilitated by heating of the parent asteroids. It is likely that such raised temperature conditions caused the same reaction mechanisms that resulted in hydrogen loss in both IOM and SOM.

This hypothesis can be tested by comparing the degree of H-loss between the IOM and the SOM. Averaged H/C ratios of SOM and that of IOM decrease with increasing alteration degree, TL11i (the most altered) < TL11h < TL5b (the least altered). By using TL5b as a reference, we semi-quantitatively estimated the H-loss in TL11h and TL11i SOM. The estimated H-loss from SOM was approximately 10-30%, while that from IOM it was 18% and 30% for TL11h and TL11i, respectively (S10). And, this consistent H-loss caused by alteration in both SOM and IOM indicates that the primordial organics compounds matured during the alteration regardless of their size. As a consequence of oxidation in both SOM and IOM, organics released hydrogen to the environment, likely affecting subsequent chemical reactions on the asteroids. Hydrogen is also known to be produced by metal oxidation during the aqueous alteration on the chondrite parent bodies. Therefore, at least two distinct mechanisms can be major contributors of $H_2$ in chondrite parent bodies. The hydrogen degassing has been considered to an essential step to explain H-isotope heterogeneities found in chondrite components. Alexander et al. 2010 suggested that the Rayleigh-type fractionation, caused by $H_2$ degassing, can create D-rich water by the equilibrium between remaining deuterium-rich $H_2$ and water. Subsequently, the D-rich water-organic reactions may have produced D-rich organics. Moreover, $H_2$ in the environment has been considered to react with oxide metals. $H_2$ produced from organics could have reacted with surrounded minerals including dissolved carbonates (Zolotov et al. 2006 and Guo and Eiler 2007). Interestingly, the total abundance of carbonate minerals in the three Tagish Lake lithologies also correlates with the degree of aqueous alterations 8, 5, and 4 vol% for TL5b (the least altered), TL11h, and TL11i (the most altered), respectively. Further C-isotope studies in the individual carbonate grains in the three samples may reveal details concerning the organic and inorganic carbon chemistry and the timing relative to the $H_2$ producing events.

In addition to the loss of hydrogen, the alteration produced an increase in small molecules (shift to low mass in Fig 2. and supplemental figures). Previously observed MCA in Tagish Lake meteorite are abundant, 100 ppm for bulk rock (Pizzarello et al. 2001). The minimum estimate for individual lithologies are 500 ppm for 5b and 11h, 300 ppm for 11i (Hilts et al. 2014). Previous studies showed that the concentration distribution of MCA $CH_2$ families in the Tagish Lake meteorite changed with increasing alteration (Hilts et al. 2014, Fig 6b). The abundance of MCA with carbon number ranging from 4 to 10 in TL5b became low relative to formic acid (C1) and acetic acid (C2) in TL11i. The diminished diversity in size by alteration is consistent with the overall trend seen in Fig 2. The distribution becomes sharper when compared to the less altered broad spectrum due to a decrease for large carbon number molecules and an increase at small carbon numbers. Thus, our study agrees with the Hilts et al. (2014) results and extends the observations to ~4200-5000 chemical formulae.

Finally, our results can be utilized for assessing the SOM alteration degree in samples that have experienced aqueous alteration on their various chondritic components. For example, Naraoka and Hashiguchi (2018) found that alkylpiperidines are more abundant in Yamato 002540 (CR) than that found in Murray (CM2), while alkylpyridines concentrations are high in Murray relative to Yamato 002540. By investigating their mass distributions, one can estimate the degree of alterations that had taken place on their parent bodies. Furthermore, studies of samples such as those from the asteroid missions to Ryugu and Bennu that were brought back by the JAXA Hayabusa2 and or will be brought back the NASA OSIRIS-Rex spacecraft, respectively, can be impacted by our findings. These asteroids consist of carbonaceous chondrite material that has undergone parent body alteration. For example, one can assess the bio-related molecular synthesis found due to parent body alteration by calculating the residual from the SZ model for homologous series as used in this study.

## 5. CONCLUSION

We studied the mass-distribution of SOM extracted from three distinct Tagish Lake meteorite fragments. We found that 1) the mass-distributions of the SOM found in the primordial fragments can be explained by using the Schulz-Zimm model; 2) the mass distribution pattern becomes asymmetric and sharper with increasing degrees of aqueous alteration; and 3) the carbon atoms in these compounds became more oxidized with increasing degrees of aqueous alteration. These observations imply that the complex nature of the primordial SOM 1) was established before severe alteration on the asteroid and that mechanisms accounting for the alteration can be explained by chain-growth polymerization; 2) was possibly established in the solar nebula; 3) was simplified on the asteroid; and 4) matured together with IOM. Therefore, it appears that the reaction pathways that have been suggested for explaining complex SOM on asteroids are not necessarily correct since the complexities seem to be preexisting even though aqueous reactions were necessary to synthesize specific biotically relevant compounds such as amino acids (Koga et al. 2021), ribose (Furukawa et al. 2019), or RNA (Cafferty and Hud 2014).

Furthermore, one can identify which organic chemical formulae were preferentially synthesized through metamorphism and alteration on asteroids relative to other solar nebula materials by measuring deviations from the SZ model.

## Acknowledgments

We thank Dr. Lee Frost Bargatze for constructive criticism of the manuscript. This work is supported by the French National Research Agency in the framework of the Investissements d'Avenir program (ANR-15-IDEX-02), through the funding of the "Origin of Life" project of the Univ. Grenoble-Alpes. PB acknowledges funding from the European Research Council under the H2020 framework program/ERC grant agreement no. 771691 (Solarys)

## Appendix

### 1. Orbitrap Measurement Methods

The SOM extracted from the Tagish Lake meteorite powder was measured by the high-resolution mass spectra, Thermo LTQ Orbitrap XL instrument coupled with an Electrospray ionization (ESI) source in the m/z range of 150—800 at the University of Grenoble Alps. The positive ions were analyzed with a resolving power of m/Δm ~ 100000 at m/z = 400u. For the analyses, the direct infusion technique was used. The SOM, together with the solvent, was injected continuously into the Orbitrap at the flow rate of 3 $\mu$l/min. The flow rate was controlled by the syringe pump attached to the instrument, and the PEEK capillary tube was used. One direct infusion analysis took approximately 30 min.

### 2. Data Analyses

After we acquired the spectra, we utilized post-processing data analysis tools, which have been developed in-house at IPAG. ATTRIBUTOR was used for main data analyses, and the following data treatment was applied using in-house code written in R language. The post-processing includes peak detection, noise rejection, and mass-drift correction. After we refined the mass spectra, we assigned the chemical formula to the detected peaks. The chemical formulae were assigned using $^{12}C$, $^{13}C$, H, $^{14}N$, and $^{16}O$. The assigned chemical formulae were filtered based on the offset from the ideal mass calculated based on the assigned chemical formula. The filter was ±1.5 ppm. In this paper, we only used the chemical formula assigned with the 12C.

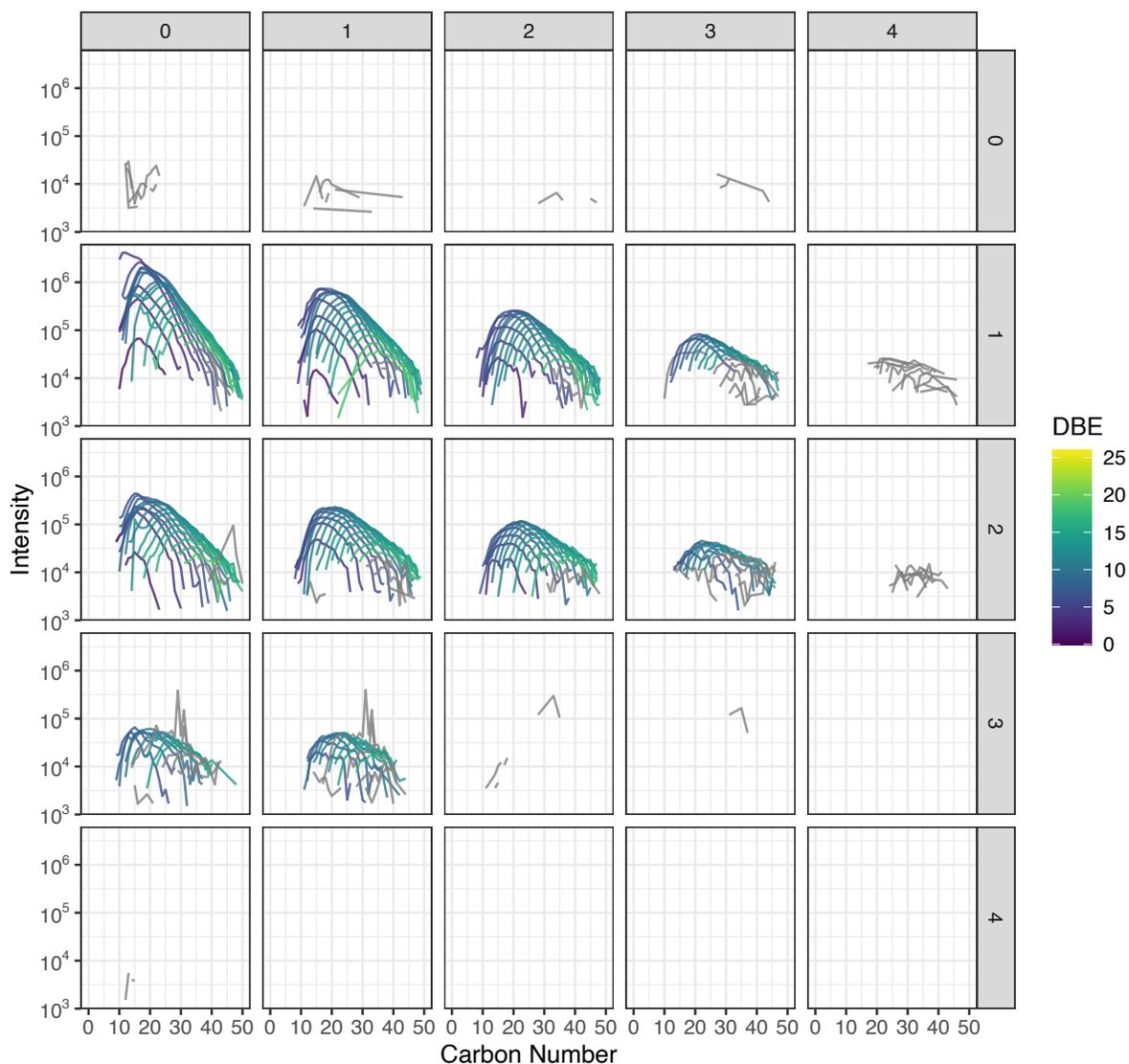

rows = numbers of N atoms and columns = numbers of O atoms

**S1:** Modified mass spectra as a function of carbon number. The results are from the samples, TL5b. We treated the stoichiometric formulae varying in $CH_2$ as a repeating unit and call them as $CH_2$ family respect to DBE values after the chemical assignment. They are treated similar to alkyl homologous compounds that identified in previous study of Murchison meteorite (Naraoka et al. 2017). The data were also organized by the number of heteroatoms, N and O coloured by DBE values. The data in grey are filtered out for the figure 2. The filtering criteria is number of peaks in a $CH_2$ family is larger than 10 and calculated entropy > 2 and the goodness of the SZ fit > 0.5.

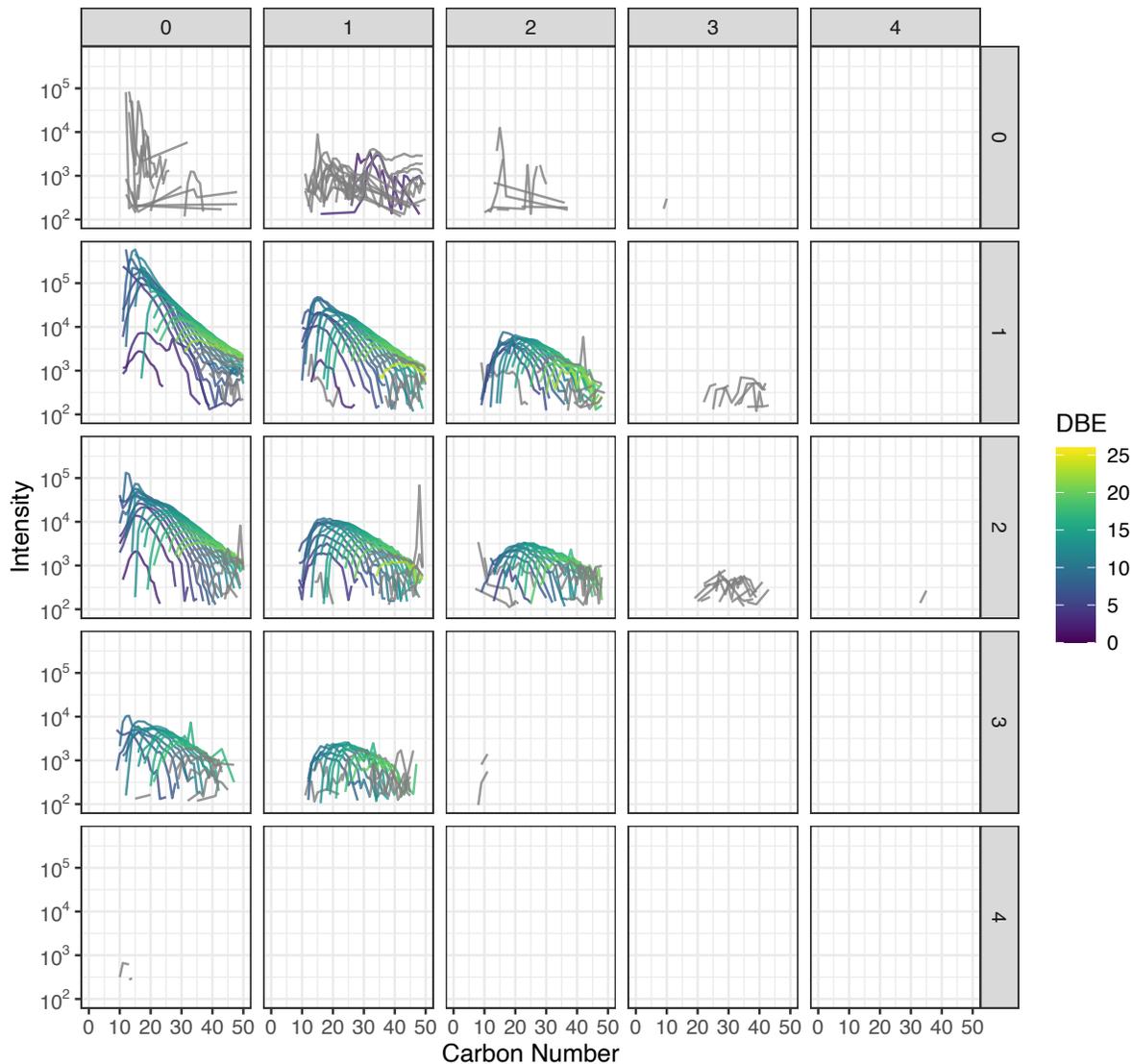

rows = numbers of N atoms and columns = numbers of O atoms

**S2**: Modified mass spectra as a function of carbon number. The results are from the samples, TL11h. We treated the stoichiometric formulae varying in $CH_2$ as a repeating unit and call them as $CH_2$ family respect to DBE values after the chemical assignment. They are treated similar to alkyl homologous compounds that identified in previous study of Murchison meteorite (Naraoka et al. 2017). The data were also organized by the number of heteroatoms, N and O coloured by DBE values. The data in grey are filtered out for the figure 2. The filtering criteria is number of peaks in a $CH_2$ family is larger than 10 and calculated entropy > 2 and the goodness of the SZ fit > 0.5.

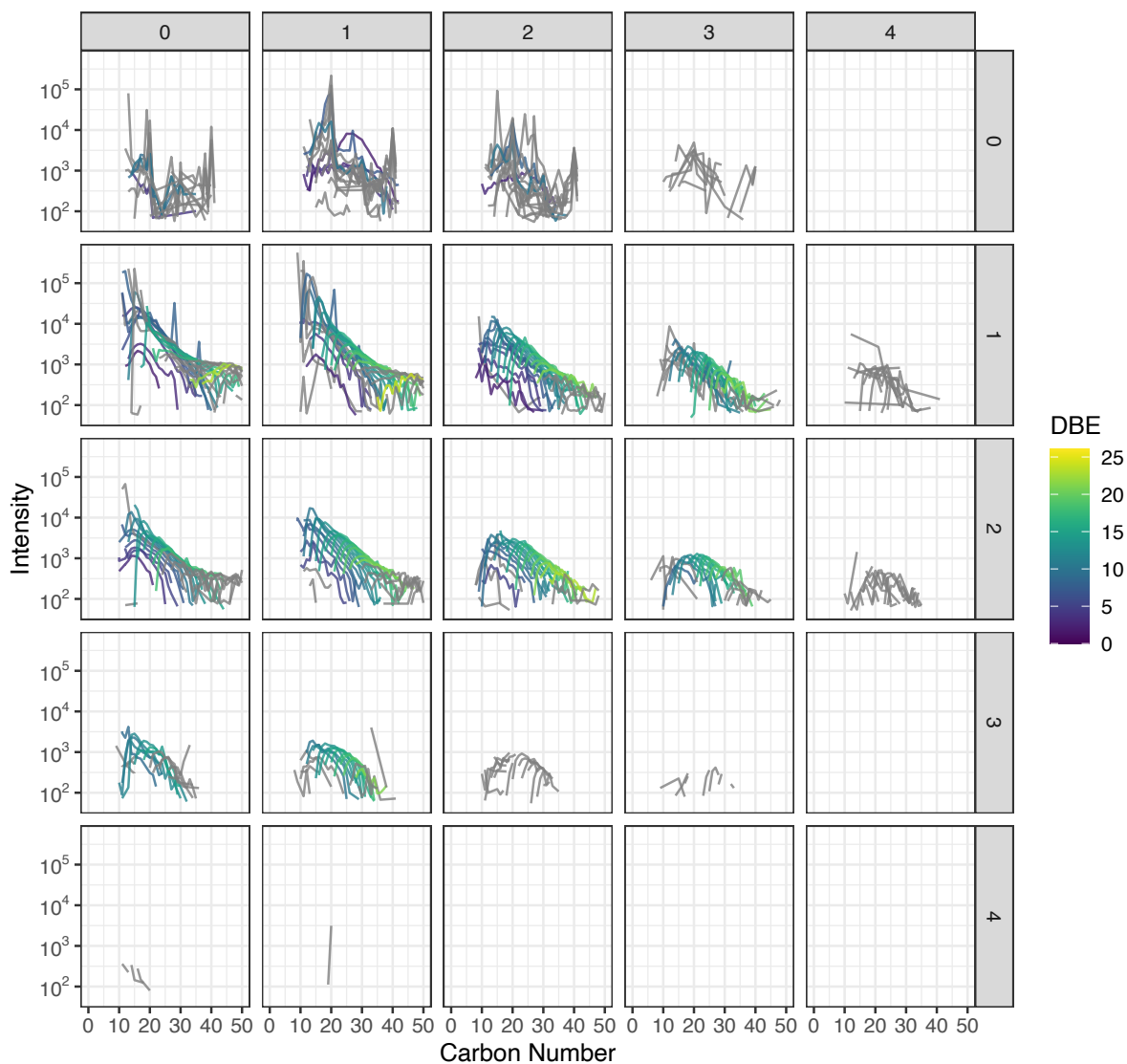

rows = numbers of N atoms and columns = numbers of O atoms

**S3:** Modified mass spectra as a function of carbon number. The results are from the samples, TL11i. We treated the stoichiometric formulae varying in $CH_2$ as a repeating unit and call them as $CH_2$ family respect to DBE values after the chemical assignment. They are treated similar to alkyl homologous compounds that identified in previous study of Murchison meteorite (Naraoka et al. 2017). The data were also organized by the number of heteroatoms, N and O coloured by DBE values. The data in grey are filtered out for the figure 2. The filtering criteria is number of peaks in a $CH_2$ family is larger than 10 and calculated entropy > 2 and the goodness of the SZ fit > 0.5.

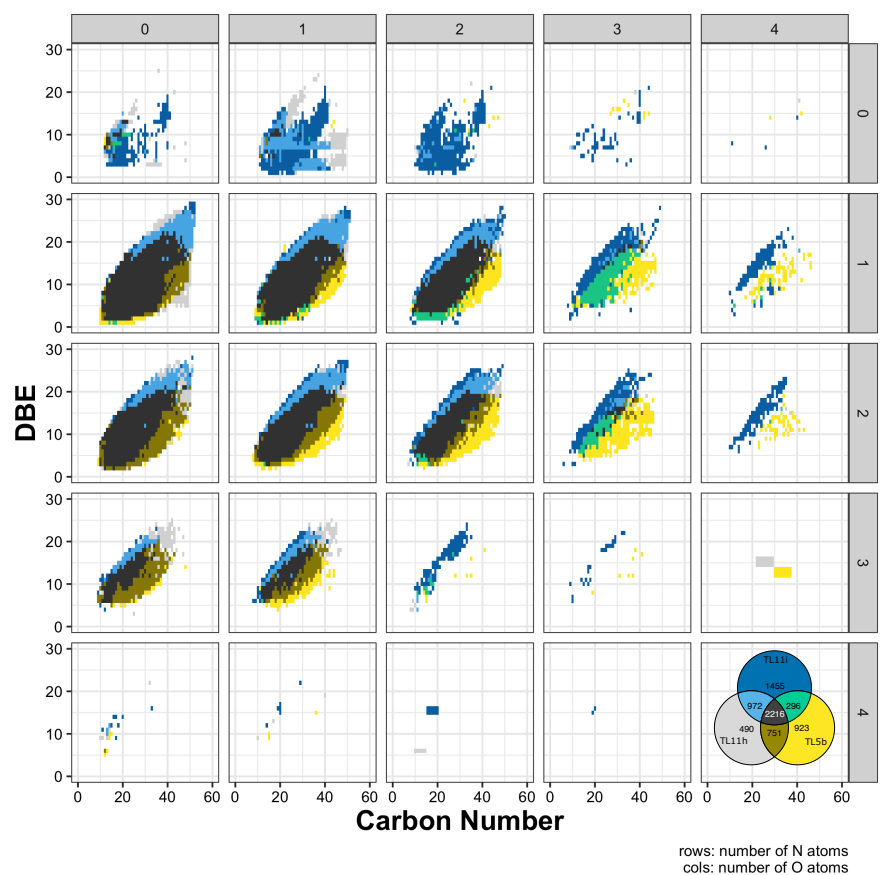

**S4:** The common and divergence of assigned chemical formulae. The individual box is organized the number of heteroatoms in the chemical formulae. One pixel indicates one chemical formula. The data were color-coded by samples (see Venn diagram). The least altered samples tend to be plotted at the bottom of the point clouds in the individual boxes (yellow). The altered samples tend to be plotted at the top of the point clouds in the individual boxes (blue). This trend indicates that the compounds became oxidized, which is either become more cyclic or forming double bonds.

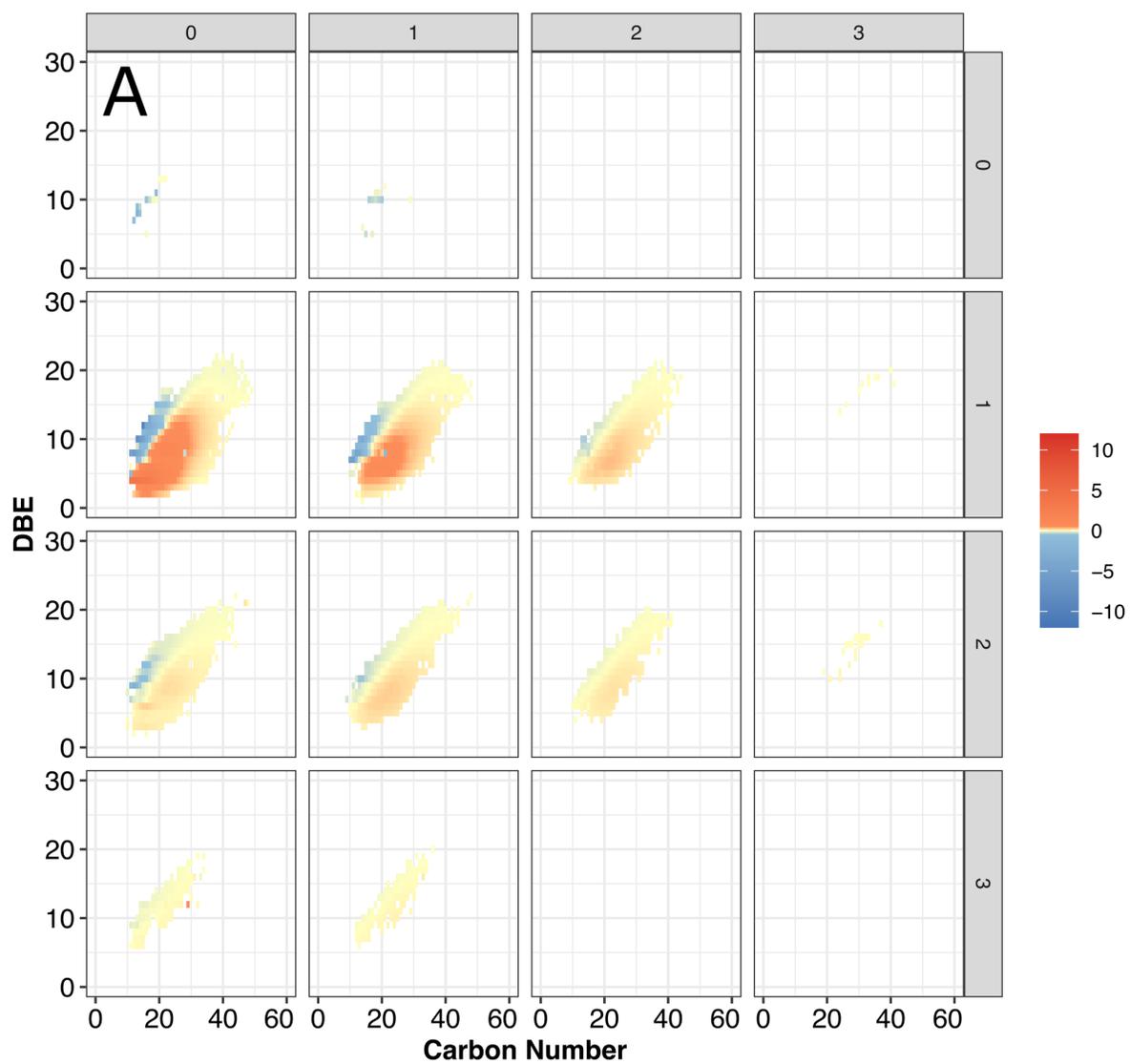

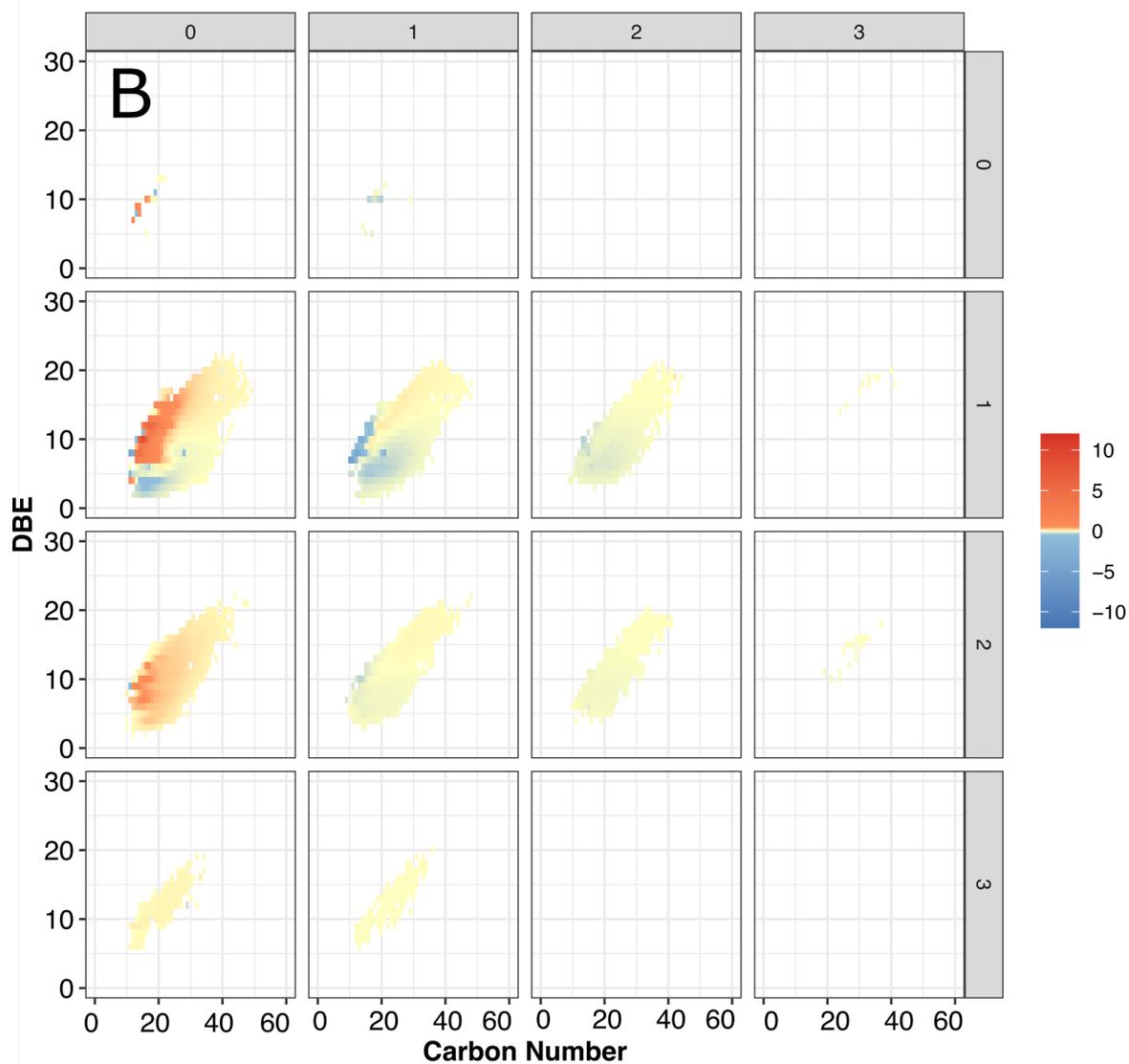

row = number of N atoms and cols = number of O atoms

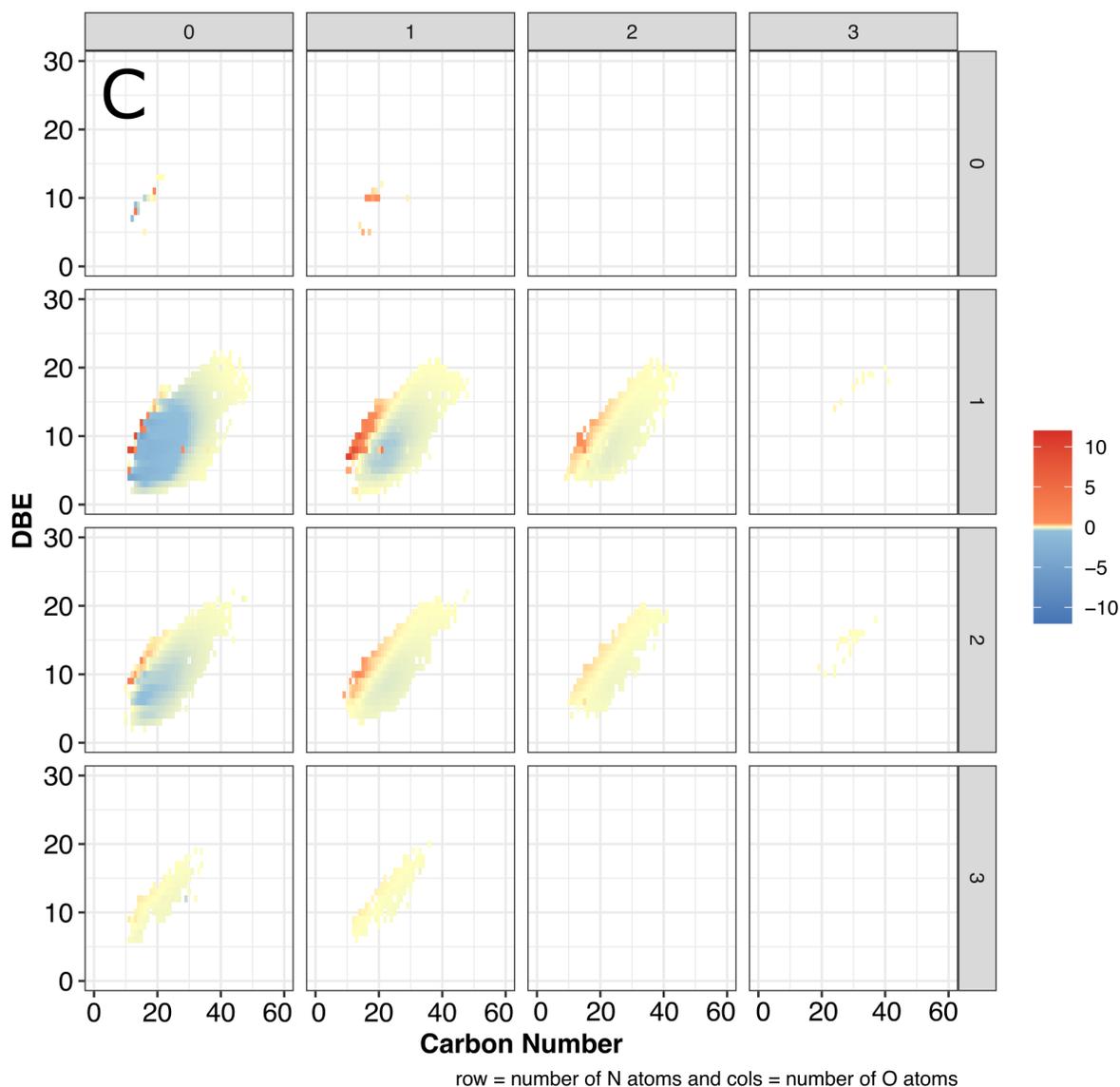

**S5:** The common chemical formulae, sample 5b, sample 11h, and sample 11i (at the center of the Venn diagram see S4). Normalized intensities among the three samples. The figure A, B, C, represent the sample 5b, 11h, 11i, respectively. The positive (red) or negative (blue) index indicates that the intensities that are higher or lower than the normalized average intensities. With an increasing degree of alteration, the larger DBE chemical formulae intensity increases.

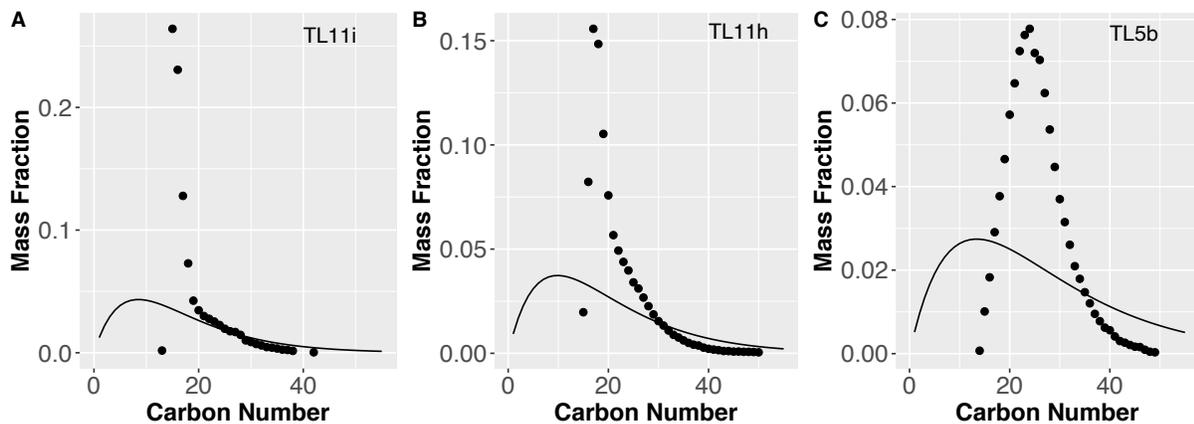

**S6:** An example CH$_2$ family, C$_{14}$H$_9$NO + (CH$_2$)$_n$ **(A–C)**. Same data are plotted in the Fig2 (A-C). Black solid line is the best fit of Anderson-Flory-Schultz (AFS) distribution to the data set. The data are not well explained by the AFS model.

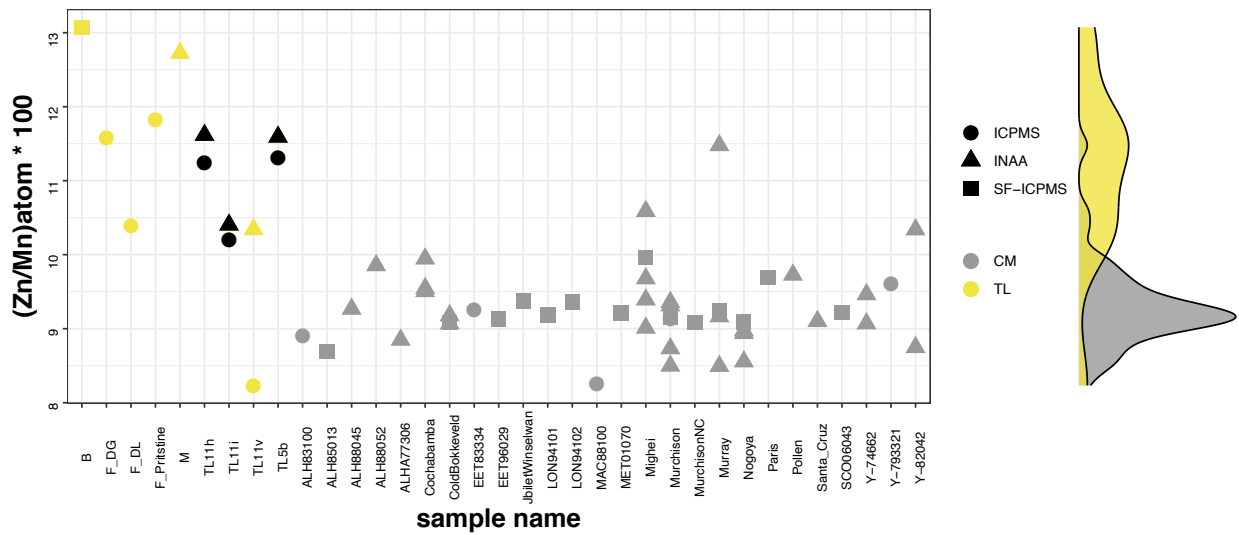

**S7:** Bulk Zn/Mn ratios of the Tagish Lake meteorite, CM group based on the previouslly reported data. The moderately volatile element, Mn, and volatile element, Zn, the ratio is one of the representations of accretionary material diversity. Assuming that the scatter within the CM is an example of a typical variety of chondrite groups, the samples used in this study TL11h, TL11i, and TL5b (black symbols) are homogeneous, while Tagish Lake lithology TL11v, for example, shows relatively large heterogeneity. That disagreement in the aliquots can indicate the contamination from the foreign clast, which was located in the Tagish Lake meteorite. On the contrary, the lithology, TL11h, TL11i, and TL5b, are closer to the average Tagish Lake meteorite value. Their duplicate analyses by the different aliquots using the distinct analytical techniques well agree with each other.

Tagish Lake meteorite data includes followings: B (Braukmüller et al. 2018); F_DG (Disturbed_G) (Friedrich et al. 2002); F_DL (Disturbed_L) (Friedrich et al. 2002); F_Pristine (Pristine) (Friedrich et al. 2002); M (Mittlefehldt 2002); TL11h, TL11i, TL11v, and TL11v (Blinova et al. 2014b). CM chondrites data are previously reported INAA and ICP-MS data (Braukmüller et al. 2018; Friedrich et al. 2002; Grady et al. 1987; Kallemeyn & Wasson 1981; Mittlefehldt 2002; Wlotzka et al. 1989).

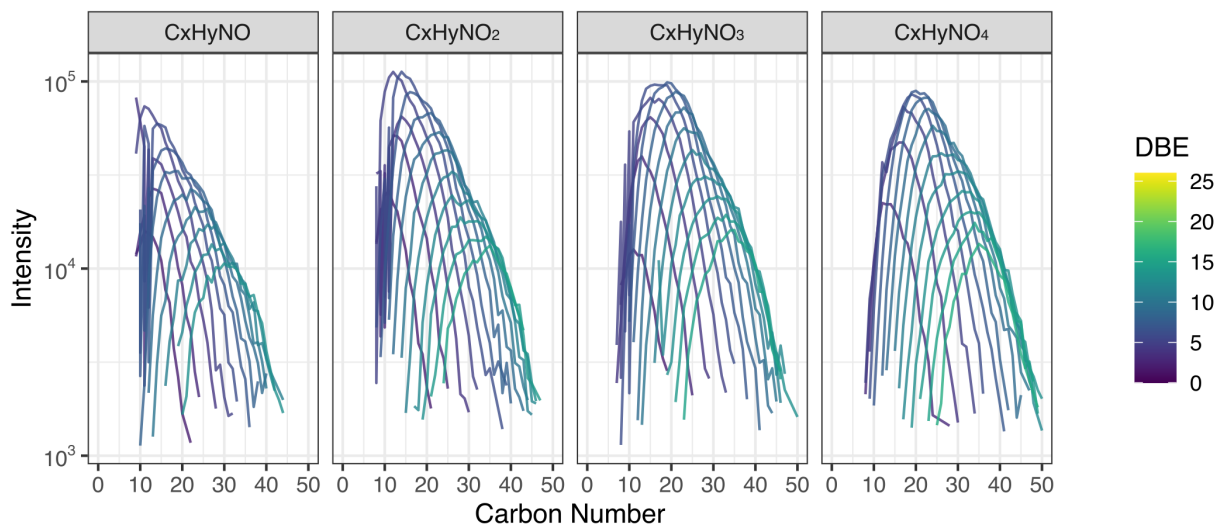

**S8**: Modified mass spectra as a function of carbon number. The results are from the samples, Nebulotron material (Zandanel 2021). We had conducted SOM extraction and analyses under the same manner with Tagish Lake meteorite (Isa et al. 2019). We treated the stoichiometric formulae varying in $CH_2$ as a repeating unit and call them as $CH_2$ family respect to DBE values after the chemical assignment. They are treated similar to alkyl homologous compounds that identified in previous study of Murchison meteorite (Naraoka et al. 2017). Molecules bearing N(1) and O(1-4) were selected and organized by the number of O. They are coloured by DBE values.

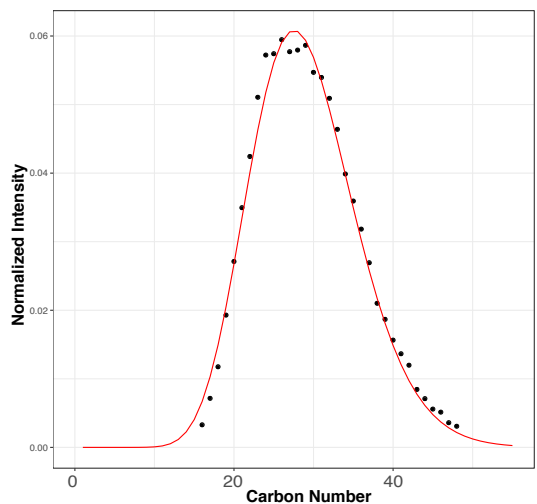

**S9**: Schulz‑Zimm (SZ) fit for the data of Nebulotoron (the data are same as S9). An example $CH_2$ family, $C_{14}H_9NO + (CH_2)_n$ is selected.

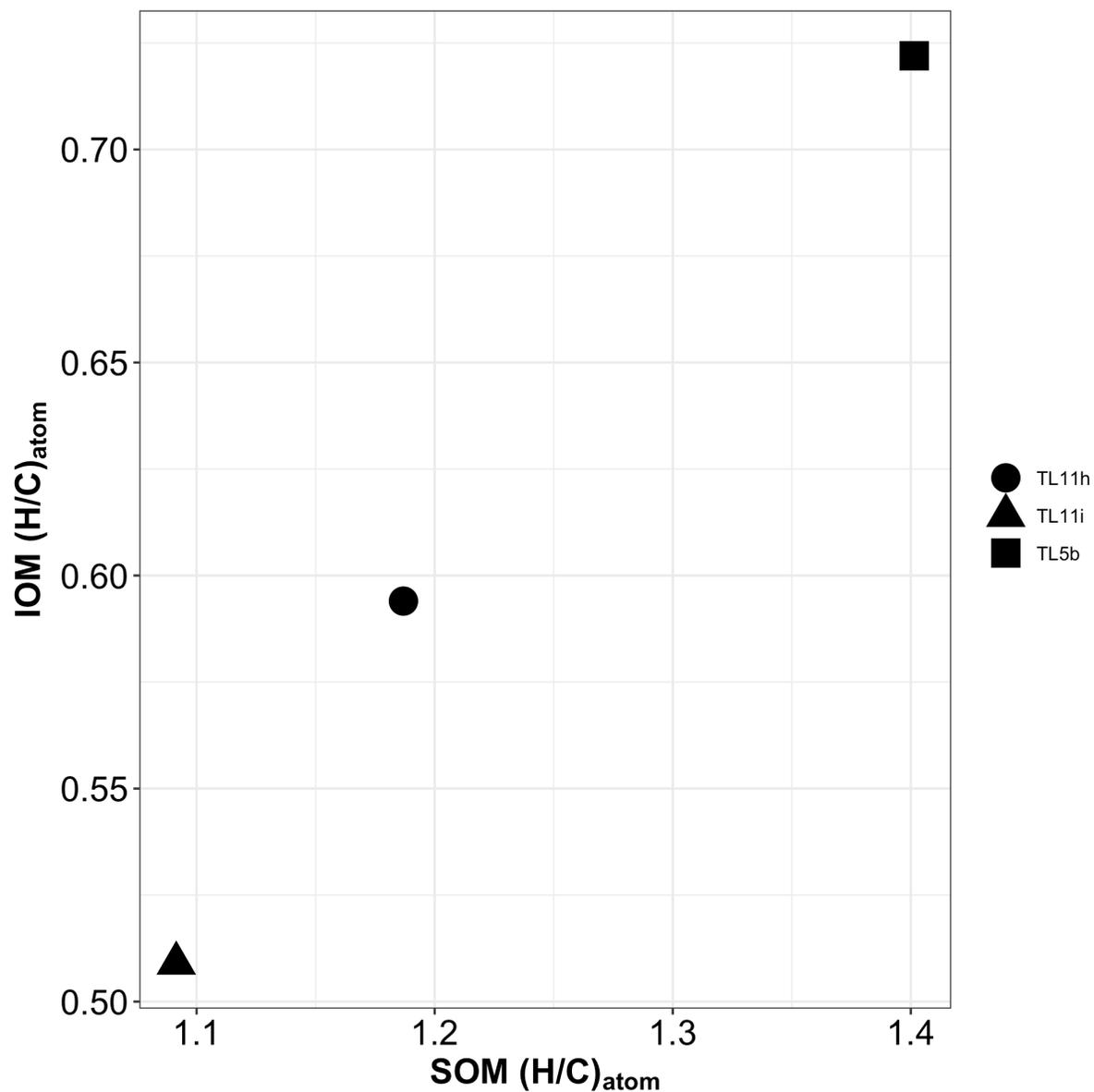

**S10:** The atom H/C ratio of SOM vs. atom H/C ratio of IOM. The SOM H/C ratios were calculated by averaging over the N-bearing assigned ions' relative intensity. For the bulk SOM H/C ratio, which is filled squared symbol in the figure, we used all the assigned spectra. The IOM data are from previous study (Alexander et al. 2014).

Table S1. Bulk rock mineralogy and componets of Tagish Lake samples

|  | 5b | 11h | 11i |
|---|---|---|---|
|  | wt % | wt % | wt % |
| Forsterite | 39 | 15 | 19 |
| Magnetite | 30 | 18 | 19 |
| Pyrrhotite | 8 | 8 | 4 |
| Enstatite | 3 | n/d | n/d |
| Dolomite | 2 | n/d | n/d |
| Siderite | 10 | 10 | 7 |
| Calcite | 2 | 0 | 1 |
| Clinochlore | n/d | 48 | 48 |
| Amorphous | 7.6 | 1.7 | 2.6 |
|  | vol% | vol% | vol% |
| Chondrules | 30 | 25 | 5 |
| Magnetite | 18 | 15 | 15 |
| Sulfides | 7 | 4 | 2 |
| Isolated silicate | 10 | 1 | 5 |
| Carbonate | 8 | 5 | 4 |
| Lithic fragments | 2 | 10 | 13 |
| Matrix | 25 | 42 | 54 |

The data from Blinova et al. 2014a

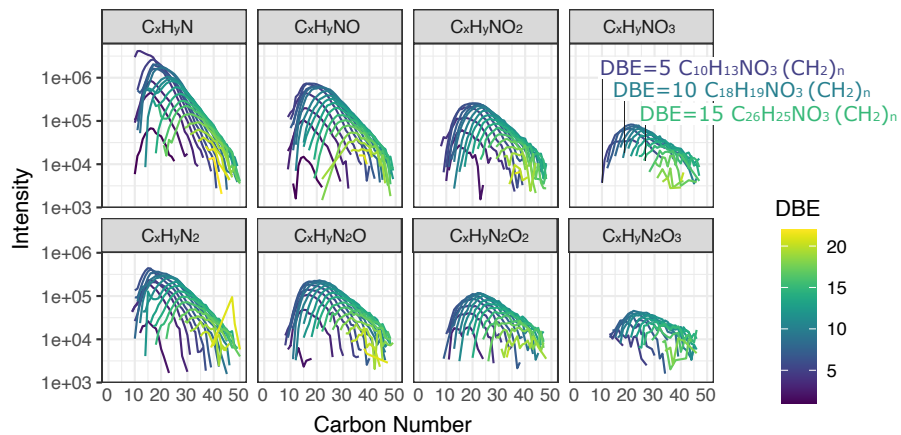

**Figure 1:** Modified mass spectra as a function of carbon number. The *m/z* that is mass spectra's x-axis is converted to carbon number. The results are from the least altered Tagish Lake samples, TL5b. Molecules bearing N(1-3) and O(0-3) were selected and the data were also organized by the number of heteroatoms, N and O. The completed data can be found in the supplement S1. DBE stands for Double Bond Equivalent and denotes the level of unsaturation.

We treated the molecular formulae varying in $CH_2$ as a repeating unit and call them as $CH_2$ family respect to DBE values after the chemical assignment. They are treated similar to alkyl homologous compounds that identified in previous study of Murchison meteorite (Naraoka et al. 2017).

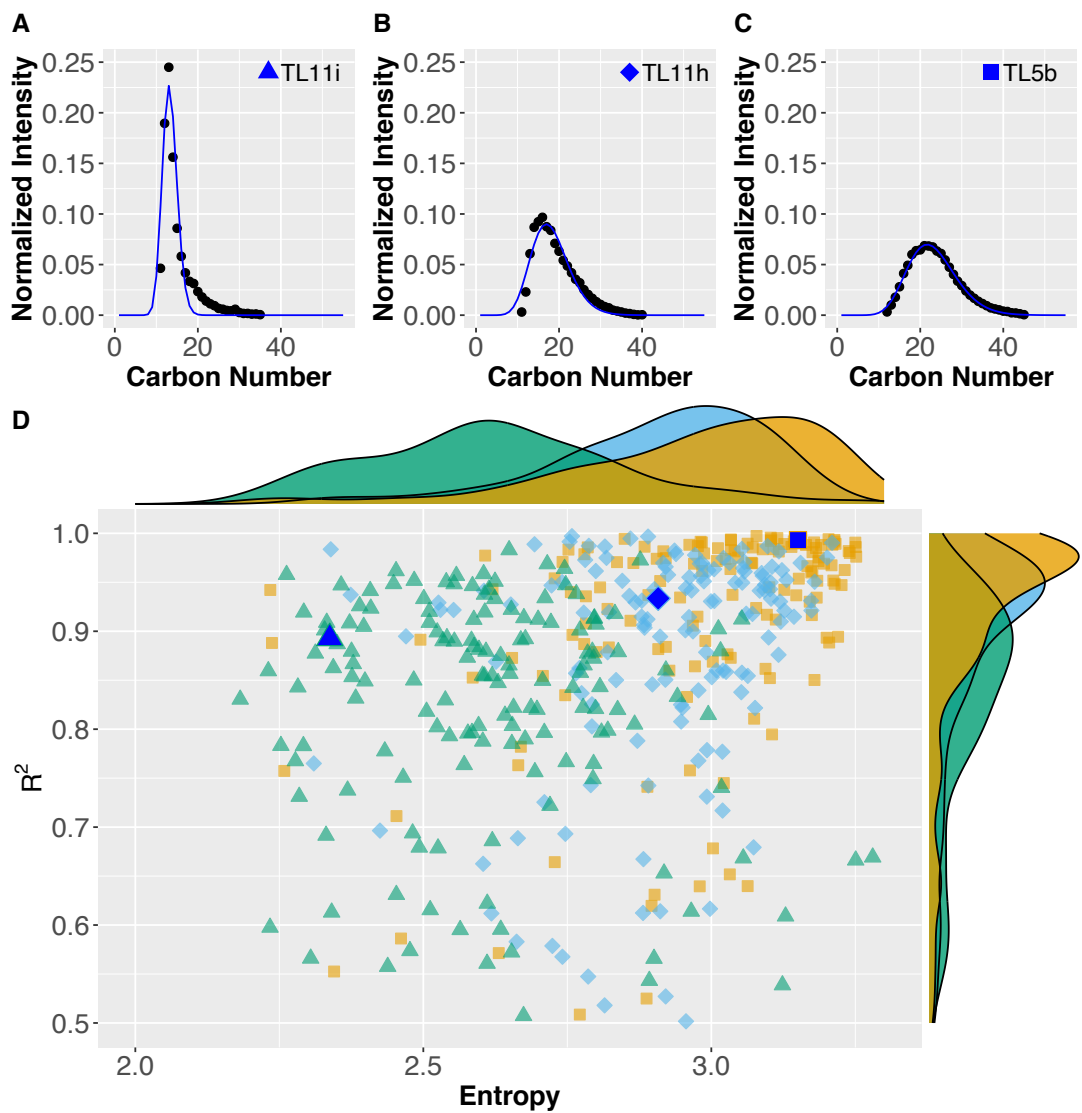

**Figure 2:** Diversity in size and Schulz-Zimm (SZ) fit for the data. An example $CH_2$ family, $C_{14}H_9NO + (CH_2)_n$ is selected **(A-C)**. The three panels **(A-C)** show data from Orbitrap analyses that are arranged in order by the degree of aqueous alteration (11i > 11h > 5b). The rightest panel **(C)** corresponds to the same data in the Fig. 1 that is presented in the first rows (N=1) of the second column (O=1) with the green line for DBE equal to 11. The blue curve in the panels **(A-C)** is a fit to the data found via least squares fitting by using a SZ distribution. The curve that can be inferred from the empirical data increasingly diverges from the SZ curves with increasing degree of alteration. The total amount of detected organic compounds in the three extracted solution is approximately 11i and 11h > 5b by the relative intensity against the strong back ground peak. Thus, the degradations from the model that we see along with the degree of aqueous alteration are not due to the low concentration of SOM in the analysed solution. The panel **(D)** is calculated entropy vs goodness of fit to the SZ distribution of an individual $CH_2$ family. Although it would be straightforward to plot the shape of all individual spectra (e.g. panel A-C), we calculated the Shannon's diversity index aka entropy $H_{\text{CH2 family}} = -\sum P_i \ln P_i$ where $P_i$ is probability: normalized intensity for a given number of carbons in a $CH_2$ family (e.g. y-axis of the panel A-C); and *i* is the number of carbons in a

$CH_2$ family (e.g. x- axis of the panel A-C). The individual density diagram at the top and right side of the figure indicates frequency density diagrams of the entropy and the good ness of the fit, respectively. The blue filled triangle, diamond, and circle symbols correspond to a $CH_2$ family, $C_{14}H_9NO + (CH_2)_n$, shown in the panel A, B, and C, respectively.

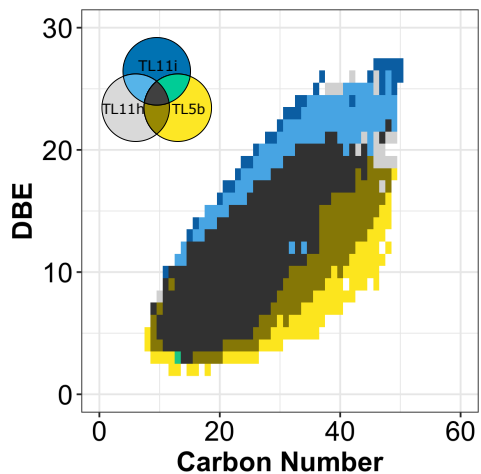

**Figure 3:** The common and divergence of assigned chemical formulae. Two N- and one O-bearing formulae were selected (completed dataset can be found in the supplement S1). One pixel indicates one chemical formula. The data were color-coded by samples (see Venn diagram). The least altered samples tend to be plotted at the bottom of the point clouds in the individual boxes (yellow). The most altered samples tend to be plotted at the top of the point clouds in the individual boxes (blue). This trend indicates that the compounds became oxidized, which is either become more cyclic or forming double bonds by losing H from the molecules.